\pdfoutput=1
\documentclass{article}

\PassOptionsToPackage{numbers, compress}{natbib}

    \usepackage[preprint]{neurips_2023}

\usepackage[utf8]{inputenc} 
\usepackage[T1]{fontenc}    
\usepackage{hyperref}       
\usepackage{url}            
\usepackage{booktabs}       
\usepackage{amsfonts}       
\usepackage{nicefrac}       
\usepackage{microtype}      
\usepackage{xcolor}         
\usepackage{hyperref}
\hypersetup{
    colorlinks,
    linkcolor={red!50!black},
    citecolor={blue!50!black},
    urlcolor={blue!80!black}
}

\usepackage{tcolorbox}

\iffalse
    \newcommand{\cz}[1]{\textcolor{blue}{}}
\else
    \newcommand{\cz}[1]{\textcolor{blue}{\bf\small [CZ: #1]}}
\fi 

\iffalse
    \newcommand{\kc}[1]{\textcolor{}{}}
\else
    \newcommand{\kc}[1]{\textcolor{}{\bf\small [KC: #1]}}
\fi 

\title{AI for Mathematics: A Cognitive Science Perspective}

\author{
    Cedegao E. Zhang* \\
    MIT BCS \\
    \texttt{cedzhang@mit.edu} \\
    \And
    Katherine M. Collins* \\
    University of Cambridge \\
    \texttt{kmc61@cam.ac.uk} \\
    \And
    Adrian Weller \\
    University of Cambridge \\ Alan Turing Institute \\
    \texttt{aw665@cam.ac.uk}
    \And
    Joshua B. Tenenbaum \\
    MIT BCS \\
    \texttt{jbt@mit.edu}
}

\begin{document}

\maketitle

\def\thefootnote{*}\footnotetext{These authors contributed equally to this work.}\def\thefootnote{\arabic{footnote}}

\begin{abstract}

Mathematics is one of the most powerful conceptual systems developed and used by the human species. Dreams of automated mathematicians have a storied history in artificial intelligence (AI). Rapid progress in AI, particularly propelled by advances in large language models (LLMs), has sparked renewed, widespread interest in building such systems. In this work, we reflect on these goals from a \textit{cognitive science} perspective. We call attention to several classical and ongoing research directions from cognitive science, which we believe are valuable for AI practitioners to consider when seeking to build truly human (or superhuman)-level mathematical systems. We close with open discussions and questions that we believe necessitate a multi-disciplinary perspective---cognitive scientists working in tandem with AI researchers and mathematicians---as we move toward better mathematical AI systems which not only help us push the frontier of the mathematics, but also offer glimpses into how we as humans are even capable of such great cognitive feats. 
\end{abstract}

\section{Introduction}

Building computational systems that understand and practice mathematics at the level of human mathematicians has been a long-standing aspiration of artificial intelligence (AI)~\citep{turing1948intelligent, newell1956logic,  newell1957empirical, wang1960toward, davis1962machine, tarski1969truth, lenat1976am, bledsoe1977nonresolution, bundy1983computer, bundy1988use, bundy1993rippling, schulz2002e, moura2008z3, ganesalingam2013fully, davies2021advancing}. The rise of large language models (LLMs) has sparked imaginations that we are closer than ever to attaining, or surpassing, human-level performance on a range of tasks~\citep{brown2020language, openai2023gpt4, bubeck2023sparks}. Yet, simultaneously, something seems amiss: despite these models achieving tremendous performance in many realms of human expertise (e.g., medicine, law, creative writing), the performance of these models on \textit{mathematics} specifically lags behind~\citep{collins2023evaluating, frieder2023mathematical, wu2023empirical, dziri2023faith}. There are many efforts to improve the mathematical problem-solving capabilities of LLMs, such as adjusting the training data and feedback strategies~\cite{lewkowycz2022solving, luo2023wizardmath, lightman2023let, yu2023metamath, gunasekar2023textbooks}, equipping models with expanded background knowledge at inference-time~\citep{welleck2022naturalprover}, or composing LLMs with existing computational mathematics systems~\citep{jiang2022draft, jiang2022thor, davis2023testing, first2023baldur, he-yueya2023solving}. Recent efforts to build in principles from cognitive science, e.g., the importance of learning abstractions, have also seen success~\cite{li2022lemma, poesia2023peano}; however, we believe that the broader AI-mathematics community still has much to draw from cognitive science: in the questions we ask and the methods by which we approach such challenges. 

First, we believe it is essential to reflect on what goals we are even trying to pursue. What does it \textit{mean} for AI systems to excel at mathematics \textit{at or beyond} a human level? Is simply excelling at a suite of standard benchmark datasets sufficient? While there is no doubt value in benchmarks to spur progress, humans---and human mathematicians---are capable of so much more than what can be captured in a static benchmark. We are capable of \textit{intuitions} and \textit{judgments}~\citep{dehaene2009origins}, of reasoning about the world \textit{as world}~\citep{smith2019promise}, of seeking deeper explanations and understandings of results~\citep{mancosu2001mathematical}, of flexibly developing new problem solving tactics and not just solving new problems, but \textit{posing} them too~\citep{silver1996analysis, schulz2012finding}. How then should we proceed to develop human-level AI mathematicians? In the rest of this position paper, we argue that perspectives from cognitive science have a lot to offer in this new age of LLMs. Cognitive scientists, AI researchers, and mathematicians can productively contribute together to this vision towards growing flexible, automated mathematicians that help us push the frontiers of mathematical knowledge and reflect back on how we are even capable of remarkable achievements of mathematical cognition~\citep{mcclelland2022capturing}.

\section{Looking to cognitive science}

We now call attention to several classical and active research directions within cognitive science which we believe hold value for those building mathematical AI systems. 

\subsection{Sample-efficient learning}

One of the hallmarks of human cognition is our ability to learn new concepts, knowledge, and problem-solving strategies, from little data~\cite{carey2009origin, tenenbaum2011how, lake2015human, gopnik2003theory, kaminski2008advantage, mitchell2021abstraction}. In mathematics, data paucity is a particular conundrum, e.g., it is costly and difficult to obtain high-quality data on advanced topics, and few texts may exist on the cutting-edge or more obscure branches of mathematics. On the other hand, human mathematicians, from early learners to expert-level mathematicians, do not need millions of examples to learn mathematical concepts and problem-solving strategies. Yet, even though the rote \textit{number} of examples developing mathematicians may be exposed to is small, that does not mean that a concept is grasped \textit{immediately} upon exposure. It make take a human multiple encounters with an example, extended time sitting and thinking---squeezing out a tremendous amount from a handful of examples, e.g., through active engagement like self-explanation (see below)---to master a concept or strategy, after which such knowledge can be readily generalized to new situations~\citep{lake2015human, zhu1987learning, dehaene1992varieties, alcock2015investigating, lester2016can}. 

\subsection{Concepts, representations, and world models}

It is inspiring to reflect on the sample-efficiency of human learning. If we are to obtain or surpass such capabilities in AI systems, it is important to examine \textit{how} humans may achieve such efficiency in the first place. Towards this end, we point to the rich cognitive science literature on concepts, their representations, and how the human mind builds rich models of the world out of concepts~\cite{block1986advertisement, murphy2004big, goodman2014concepts, lake2023word}. 

In cognitive science, much research in cognitive science points to powerful inductive biases gleaned through evolution: ``core knowledge''~\citep{spelke2000core, spelke2007core}. It has been speculated that a core ``number sense''~\citep{dehaene2011number} forms the foundation upon which our mathematical prowess is built. Strong evidence points to two core number systems---for reasoning about numerosity exactly, and approximately~\citep{dehaene1992varieties, dehaene2011number, feigenson2004core}. From these core knowledge systems, we can develop \textit{concepts}~\citep{carey2009origin, goodman2014concepts, margolis2015the}. Notably in mathematics, concepts have \textit{precise definitions}, unlike other abstract concepts such as justice and knowledge or everyday concepts like chair. At the same time, mathematicians think about concepts more than in terms of definitions; they can give examples and counterexamples, draw out relationships between concepts, and so on---this type of conceptual richness is compatible with the psychological theory of conceptual-role semantics~\citep{block1986advertisement}.

So, what are the \textit{form(s)} of these conceptual representations? Contemporary cognitive science has provided strong evidence for that conceptual representations may be modeled by ``languages of thought'' ~\citep{fodor1975language, piantadosi2016logical, dehaene2022symbols, wong2023word}, which in mathematics, may be built over core geometric primitives~\citep{sable-meyer2022language}. Closely linked with ``languages of thought'' is the notion of a ``world model''. In AI, many have highlighted the importance of world models, although researchers disagree about how to build such models within AI systems~\citep{li2022emergent, matsuo2022deep, mialon2023augmented, wong2023word}. It is generally accepted that a world model should support simulation of possibilities, causal and counterfactual reasoning, and calibrated judgements about belief and truth~\citep{tenenbaum2011how, matsuo2022deep, mialon2023augmented, tenenbaum2006theory, ullman2020bayesian, johnson-laird1989mental, rule2020child, lake2017building, pearl2018book}. We hypothesize that the intuitions that mathematicians acquire over years of practice can be seen as forming world models of the mathematical universe. Here, we use the famous ``$\mathsf{P} = \mathsf{NP}$?'' problem as an illustrative example. Most people believe that $\mathsf{P} \neq \mathsf{NP}$. It seems that much evidence of such strong beliefs over an unproven statement comes from simulating what would happen if $\mathsf{P} = \mathsf{NP}$ or $\mathsf{P} \neq \mathsf{NP}$. If the former is true, many counter-intuitive consequence would follow, whereas we would not need to heavily adjust our other beliefs about computation if the latter is true~\citep{quine1951two, aaronson2016pnp, gowers2023mathbelief}. This kind of simulation and argumentation, we suggest, may be powered by world models. 

\subsection{Goals, planning, and agency}

Today, the dominant paradigm for large language models is a passive one: a (very large) training corpus is provided to the model, and the model optimizes some given objective function~\citep{radford2019language, brown2020language}. At inference-time, a model is presented with a problem (e.g., a translation or reasoning task) and tries to make good predictions. However, this is not how humans think about or perform problem-solving. Humans are planning agents with goals spanning across different communities and timescales~\citep{bratman1987intention, tomasello2022evolution}. When planning to achieve a goal, we can flexibly divide a task into sub-goals, form and leverage simplified abstract representations to inform planning, and replan~\citep{newell1972human, tomov2020discovery, ho2022people, ho2022planning, correa2023humans}. Planning is crucial to success in mathematical reasoning. Consider when a teacher gives a student a problem to solve; the student needs to generate sub-goals and come up with strategies, such as looking up definitions, consider examples, examine different cases, or simply look for help. Moreover, mathematical cognition is not just about planning for set goals, but \textit{inventing} new goals, problems, and concepts \citep{silver1996analysis, schulz2012finding, mcclelland2022capturing, lester2016can}. How do some mathematicians \textit{form the goal} of inventing new mathematics, and how do they achieve it? Engineering and scientific insights on these questions---drawing on cognitive science, AI, and mathematics---may drive a huge leap forward towards creative AI mathematicians.

\subsection{Cognitive limitations and resource-rationality}

However, humans are far imperfect planners, and they may fail to execute the plans we do embark upon. Mathematicians may become wedded to a particular proof strategy only to realize it was misguided and need to backtrack, or worse, could fall prey to functional fixedness~\citep{adamson1952functional, ho2023rational} and the sunk cost fallacy~\citep{arkes1985psychology}. Such instances put a damper in the notion that humans are rational reasoners~\citep{shafir2002rationality}. Cognitive scientists here too have developed rich frameworks to reconcile such challenges. Rather, we may be viewed as rational \textit{in light of resource constraints}, i.e., ``resource-rational''~\citep{gershman2015computational, lieder2020resource, griffiths2020understanding, icard2023resource}. This notion finds particular importance when thinking about humans and AI systems. Fundamentally, humans and computational systems have different resource limitations: computers are able to make calculations extremely fast, are not constrained to the same limitations on working memory, and do not succumb to daily inevitable fatigue that we humans do. When building mathematical AI systems then, it is prudent to question whether we should be designing AI systems to mimic human resource constraints~\citep{griffiths2020understanding}. If trying to build a computational ``thought partner'' to complement humans and enable us to explore greater mathematical depths than we have so far, for instance, by making more calculations and proposing possible new patterns in troves of data~\citep{davies2021advancing}, then perhaps we do not want to curtail a model's resources. However, one could argue that perhaps, such resource limitations are not a failing, but rather an \textit{advantage}: for instance, empowering us to judiciously \textit{select} which problems to solve in the first place. Indeed, mathematics communities (generally) do not waste too much time on problems that people believe to be out of reach. Studying under what settings resource limitations on mathematical cognition are advantageous, and when they are not, is a ripe space for collaboration across cognitive science, mathematics, and AI, particularly when thinking about making sensible use of limited resources even present in large-scale AI systems~\citep{hoffmann2022training, schwartz2020green}. 

\subsection{Communication and explanation} 

We close our tour of cognitive science insights to spark the imaginations of those seeking to build mathematical AI by reiterating that mathematics is a \textit{group activity} consisting of communities, and development of knowledge in any intellectual community depends on effective \textit{communication}. We argue that a cognitive perspective on communication is valuable for the math-AI community for two core reasons. First, the \textit{output} of our communication amongst each other forms the bedrock of the data used to train LLMs. Second, insights from cognitive science reveal that communication can spur learning for the communicator~\citep{lombrozo2006structure}. We start by reflecting on the latter.

Ample evidence in cognitive science reveals the power of self-explanation for improving learning and generalization~\citep{lombrozo2006structure, chi1989self, chi1994eliciting, williams2010role, bonawitz2012children, hodds2014self, rittle-johnson2017developing}. Explanations can help the explainer identify abstractions to inform induction~\citep{lombrozo2006structure, williams2010role} and reveal gaps in one's own knowledge~\citep{chi1989self}, motivating information-seeking to resolve such gaps~\citep{bonawitz2012children}. At first glance, recent LLM research such as chain-of-thought-prompting~\citep{wei2022chain}, ``self-taught reasoning''~\citep{zelikman2022star}, ``self-reflection''~\citep{shinn2023reflexion} could be viewed as self-explanation to improve reasoning, but we encourage ruminating on the cognitive underpinnings. In fact, we argue that these are \textit{not} instances of self-explanation in the way that humans self-explain. For humans, self-explanation is something that we \textit{want} to do, because understanding is intrinsically valuable~\citep{williams2010role, lombrozo2011instrumental}. Thus, it is desirable to not just have new prompting strategies leveraging explanations, but systems designed with explanations at their core.

And what about communication to others? We externalize many of our inner thoughts, whether that be writing out the steps of a new proof, drawing diagrams to convey a concept, or debating with a friend what the largest possible number is. These externalized thoughts and interactions increasingly form the bedrock of training data for AI systems. Nonetheless, humans do not communicate \textit{all} of our inner thoughts; rather, we communicate what we believe is essential to convey~\citep{grice1975logic}---often requiring the listener to make inferences about what the communicator \textit{intends} to communicate (which may differ from what they \textit{actually} produced)~\citep{levinson1983pragmatics, sperber1995relevance, goodman2016pragmatic}. Such communication frameworks may be important for building mathematical AI systems that can adequately ``read between the lines'' in the data available and recognize that when providing mathematical assistance to humans, humans \textit{are capable} of such inferences (e.g., we do not always require overly verbose responses and in fact may find it less helpful in mathematics~\cite{collins2023evaluating}). 

\section{Concluding remarks}

\paragraph{Catalyzing community cross-talk} As we highlight, the cognitive science community has been studying topics deeply relevant to mathematical AI. We hope our piece helps further expose AI practitioners and mathematicians to what we believe are valuable terminology and conceptual structures from cognitive science. Cognitive scientists too can sharpen our theories from further exchanges across communities; we lay out a few strategies to facilitate such conversations. First, accessibility of higher-level mathematics is perhaps one of the most pernicious barriers to effective collaboration across cognitive scientists and AI practitioners in the space. Convenings designed to engage not just the AI and mathematics community but also cognitive scientists would aid in building a shared vocabulary across these communities. Second, there is a need for improved \textit{research tools} to empower the study of mathematics across our communities. For more than a decade, cognitive scientists and AI practitioners alike have benefited enormously from crowdsourcing platforms such as Amazon Mechanical Turk~\citep{paolacci2010running} and Prolific~\citep{palan2018prolific}. However, at present, it is hard to find targeted domain practitioners on such sites. We suggest that it would be extremely valuable for the community to discuss the idea of a ``Mechanical Turk for mathematics''; i.e., a platform where AI and cognitive scientists can post studies, questions, data gathering attempts about mathematics and mathematicians and students can participate in them. Ideally, such an effort could benefit all parties involved. Third, we note that a strong catalyst for collaboration can be a shared goal~\citep{sherif1988robbers}. We point to \textit{games} as a sensible playground which may appeal to mathematicians, cognitive scientists, and AI practitioners. Games have been ripe grounds for study in both AI~\citep{campbell2002deep, silver2016mastering, meta2022human} and cognitive science~\citep{newell1972human, gobet2004moves, tsividis2017human, allen2023using}, and as recently exposed by Poesia~\citep{poesia2023research}, aspects of mathematics itself may be cast in the language of games. We see this as a particularly exciting framing that allows us to better understand many aspects of mathematics with the help of mathematicians.

\paragraph{Looking forward} With the resurgence in interest around AI and mathematics, we emphasize the value of engaging with the cognitive science community in the quest towards more powerful automated mathematicians. Engaging across the mathematics, cognitive science, and AI communities is paramount in even defining what this quest is and where we intend to go. To close, we propose several directions of inquiry that we think the nexus of the cognitive science, AI, and mathematics communities are poised to address. For instance, advances in AI can serve as tools to help us better understand the relationship between mathematical problem-solving capabilities and the \textit{modalities} of mathematical data---language (natural and formal) alongside figures and diagrams; what makes a problem easy or hard (and how this differs across humans and AI systems); what kinds of prior knowledge, including human commonsense knowledge, is necessary to learn mathematics; and what are the computational foundations of mathematical insights. We believe that steps along these directions, taken across our communities, can not only spur the development of truly powerful AI mathematicians, but also shed light on what is so special about \textit{humans'} feats of mathematical cognition---sparking efforts to improved tailored mathematical education and push the boundaries of what we, jointly with AI systems, understand about the wonderful world of mathematics.

\section*{Acknowledgements}

We thank Timothy Gowers, Gabriel Poesia, and Roger Levy for comments on earlier drafts. We thank Noah Goodman, Raymond Wang, and Lionel Wong for discussions related to this work. We also thank Albert Jiang, Mateja Jamnik, the Human-Oriented Automated Theorem Proving System Team at Cambridge, and the Spring 2023 GPS Seminar at MIT for conversations that inspired aspects of this work. KMC acknowledges funding from the Marshall Commission and Cambridge Trust. AW acknowledges support from a Turing AI Fellowship under grant EP/V025279/1, The Alan Turing Institute, and the Leverhulme Trust via CFI. JBT acknowledges funding from AFOSR Grant \#FA9550-22-1-0387 and the MIT-IBM Watson AI Lab.

\bibliographystyle{unsrt}
\bibliography{references}

\begin{thebibliography}{100}

\bibitem{turing1948intelligent}
Alan~M. Turing.
\newblock Intelligent machinery.
\newblock Technical report, National Physical Laboratory, 1948.

\bibitem{newell1956logic}
Allen Newell and Herbert~A. Simon.
\newblock The logic theory machine--a complex information processing system.
\newblock {\em IRE Transactions on information theory}, 2(3):61--79, 1956.

\bibitem{newell1957empirical}
A.~Newell, J.~C. Shaw, and H.~A. Simon.
\newblock Empirical explorations of the logic theory machine: A case study in heuristic.
\newblock In {\em Papers Presented at the February 26-28, 1957, Western Joint Computer Conference: Techniques for Reliability}, IRE-AIEE-ACM '57 (Western), page 218–230, New York, NY, USA, 1957. Association for Computing Machinery.

\bibitem{wang1960toward}
Hao Wang.
\newblock Toward mechanical mathematics.
\newblock {\em IBM Journal of Research and Development}, 4(1):2–22, 1960.

\bibitem{davis1962machine}
Martin Davis, George Logemann, and Donald Loveland.
\newblock A machine program for theorem-proving.
\newblock {\em Communications of the ACM}, 5(7):394--397, 1962.

\bibitem{tarski1969truth}
Alfred Tarski.
\newblock Truth and proof.
\newblock {\em Scientific American}, 220(6):63--77, 1969.

\bibitem{lenat1976am}
Douglas~B. Lenat.
\newblock {\em AM: An artificial intelligence approach to discovery in mathematics as heuristic search}.
\newblock PhD thesis, Stanford University, Stanford, CA, 1976.
\newblock Ph.D. thesis.

\bibitem{bledsoe1977nonresolution}
W.~W. Bledsoe.
\newblock Non-resolution theorem proving.
\newblock {\em Artificial Intelligence}, 9:1--35, 1977.

\bibitem{bundy1983computer}
Alan Bundy.
\newblock {\em The Computer Modelling of Mathematical Reasoning}.
\newblock Academic Press, 1983.

\bibitem{bundy1988use}
Alan Bundy.
\newblock The use of explicit plans to guide inductive proofs.
\newblock In {\em 9th International Conference on Automated Deduction}, 1988.

\bibitem{bundy1993rippling}
Alan Bundy, Andrew Stevens, F.~V. Harmelen, Andrew Ireland, and Alan Smaill.
\newblock Rippling: A heuristic for guiding inductive proofs.
\newblock {\em Artificial Intelligence}, 62:185--253, 1993.

\bibitem{schulz2002e}
Stephan Schulz.
\newblock E - a brainiac theorem prover.
\newblock {\em AI Communications}, 15:111--126, 2002.

\bibitem{moura2008z3}
Leonardo~Mendonça de~Moura and Nikolaj~S. Bj{\o}rner.
\newblock Z3: An efficient smt solver.
\newblock In {\em International Conference on Tools and Algorithms for Construction and Analysis of Systems}, 2008.

\bibitem{ganesalingam2013fully}
Mohan Ganesalingam and William~Timothy Gowers.
\newblock A fully automatic problem solver with human-style output.
\newblock {\em arXiv preprint arXiv:1309.4501}, 2013.

\bibitem{davies2021advancing}
Alex Davies, Petar Veli{\v{c}}kovi{\'c}, Lars Buesing, Sam Blackwell, Daniel Zheng, Nenad Toma{\v{s}}ev, Richard Tanburn, Peter Battaglia, Charles Blundell, Andr{\'a}s Juh{\'a}sz, et~al.
\newblock Advancing mathematics by guiding human intuition with ai.
\newblock {\em Nature}, 600(7887):70--74, 2021.

\bibitem{brown2020language}
Tom Brown, Benjamin Mann, Nick Ryder, Melanie Subbiah, Jared~D Kaplan, Prafulla Dhariwal, Arvind Neelakantan, Pranav Shyam, Girish Sastry, Amanda Askell, et~al.
\newblock Language models are few-shot learners.
\newblock {\em Advances in neural information processing systems}, 33:1877--1901, 2020.

\bibitem{openai2023gpt4}
OpenAI.
\newblock Gpt-4 technical report, 2023.

\bibitem{bubeck2023sparks}
S{\'e}bastien Bubeck, Varun Chandrasekaran, Ronen Eldan, Johannes Gehrke, Eric Horvitz, Ece Kamar, Peter Lee, Yin~Tat Lee, Yuanzhi Li, Scott Lundberg, et~al.
\newblock Sparks of artificial general intelligence: Early experiments with gpt-4.
\newblock {\em arXiv preprint arXiv:2303.12712}, 2023.

\bibitem{collins2023evaluating}
Katherine~M Collins, Albert~Q Jiang, Simon Frieder, Lionel Wong, Miri Zilka, Umang Bhatt, Thomas Lukasiewicz, Yuhuai Wu, Joshua~B Tenenbaum, William Hart, et~al.
\newblock Evaluating language models for mathematics through interactions.
\newblock {\em arXiv preprint arXiv:2306.01694}, 2023.

\bibitem{frieder2023mathematical}
Simon Frieder, Luca Pinchetti, Ryan-Rhys Griffiths, Tommaso Salvatori, Thomas Lukasiewicz, Philipp~Christian Petersen, Alexis Chevalier, and Julius Berner.
\newblock Mathematical capabilities of chatgpt.
\newblock {\em arXiv preprint arXiv:2301.13867}, 2023.

\bibitem{wu2023empirical}
Yiran Wu, Feiran Jia, Shaokun Zhang, Qingyun Wu, Hangyu Li, Erkang Zhu, Yue Wang, Yin~Tat Lee, Richard Peng, and Chi Wang.
\newblock An empirical study on challenging math problem solving with gpt-4.
\newblock {\em arXiv preprint arXiv:2306.01337}, 2023.

\bibitem{dziri2023faith}
Nouha Dziri, Ximing Lu, Melanie Sclar, Xiang~Lorraine Li, Liwei Jian, Bill~Yuchen Lin, Peter West, Chandra Bhagavatula, Ronan~Le Bras, Jena~D Hwang, et~al.
\newblock Faith and fate: Limits of transformers on compositionality.
\newblock {\em arXiv preprint arXiv:2305.18654}, 2023.

\bibitem{lewkowycz2022solving}
Aitor Lewkowycz, Anders Andreassen, David Dohan, Ethan Dyer, Henryk Michalewski, Vinay Ramasesh, Ambrose Slone, Cem Anil, Imanol Schlag, Theo Gutman-Solo, et~al.
\newblock Solving quantitative reasoning problems with language models.
\newblock {\em Advances in Neural Information Processing Systems}, 35:3843--3857, 2022.

\bibitem{luo2023wizardmath}
Haipeng Luo, Qingfeng Sun, Can Xu, Pu~Zhao, Jianguang Lou, Chongyang Tao, Xiubo Geng, Qingwei Lin, Shifeng Chen, and Dongmei Zhang.
\newblock Wizardmath: Empowering mathematical reasoning for large language models via reinforced evol-instruct.
\newblock {\em arXiv preprint arXiv:2308.09583}, 2023.

\bibitem{lightman2023let}
Hunter Lightman, Vineet Kosaraju, Yura Burda, Harri Edwards, Bowen Baker, Teddy Lee, Jan Leike, John Schulman, Ilya Sutskever, and Karl Cobbe.
\newblock Let's verify step by step.
\newblock {\em arXiv preprint arXiv:2305.20050}, 2023.

\bibitem{yu2023metamath}
Longhui Yu, Weisen Jiang, Han Shi, Jincheng Yu, Zhengying Liu, Yu~Zhang, James~T. Kwok, Zhenguo Li, Adrian Weller, and Weiyang Liu.
\newblock Metamath: Bootstrap your own mathematical questions for large language models.
\newblock {\em arXiv preprint arXiv:2309.12284}, 2023.

\bibitem{gunasekar2023textbooks}
Suriya Gunasekar, Yi~Zhang, Jyoti Aneja, Caio C{\'e}sar~Teodoro Mendes, Allie Del~Giorno, Sivakanth Gopi, Mojan Javaheripi, Piero Kauffmann, Gustavo de~Rosa, Olli Saarikivi, et~al.
\newblock Textbooks are all you need.
\newblock {\em arXiv preprint arXiv:2306.11644}, 2023.

\bibitem{welleck2022naturalprover}
Sean Welleck, Jiacheng Liu, Ximing Lu, Hannaneh Hajishirzi, and Yejin Choi.
\newblock Naturalprover: Grounded mathematical proof generation with language models.
\newblock {\em Advances in Neural Information Processing Systems}, 35:4913--4927, 2022.

\bibitem{jiang2022draft}
Albert~Q. Jiang, Sean Welleck, Jin~Peng Zhou, Wenda Li, Jiacheng Liu, Mateja Jamnik, Timoth{\'e}e Lacroix, Yuhuai Wu, and Guillaume Lample.
\newblock Draft, sketch, and prove: Guiding formal theorem provers with informal proofs.
\newblock {\em arXiv preprint arXiv:2210.12283}, 2022.

\bibitem{jiang2022thor}
Albert~Q. Jiang, Wenda Li, Szymon Tworkowski, Konrad Czechowski, Tomasz Odrzyg{\'o}{\'z}d{\'z}, Piotr Mi{\l}o{\'s}, Yuhuai Wu, and Mateja Jamnik.
\newblock Thor: Wielding hammers to integrate language models and automated theorem provers.
\newblock {\em Advances in Neural Information Processing Systems}, 35:8360--8373, 2022.

\bibitem{davis2023testing}
Ernest Davis and Scott Aaronson.
\newblock Testing {GPT-4} with {Wolfram Alpha} and {Code Interpreter} plug-ins on math and science problems.
\newblock {\em arXiv preprint arXiv:2308.05713}, 2023.

\bibitem{first2023baldur}
Emily First, Markus~N. Rabe, Talia Ringer, and Yuriy Brun.
\newblock Baldur: Whole-proof generation and repair with large language models.
\newblock {\em arXiv preprint arXiv:2303.04910}, 2023.

\bibitem{he-yueya2023solving}
Joy He-Yueya, Gabriel Poesia, Rose~E. Wang, and Noah~D. Goodman.
\newblock Solving math word problems by combining language models with symbolic solvers.
\newblock {\em arXiv preprint arXiv:2304.09102}, 2023.

\bibitem{li2022lemma}
Zhening Li, Gabriel Poesia, Omar Costilla-Reyes, Noah Goodman, and Armando Solar-Lezama.
\newblock {LEMMA}: Bootstrapping high-level mathematical reasoning with learned symbolic abstractions.
\newblock {\em arXiv preprint arXiv:2211.08671}, 2022.

\bibitem{poesia2023peano}
Gabriel Poesia and Noah~D. Goodman.
\newblock Peano: learning formal mathematical reasoning.
\newblock {\em Philosophical Transactions of the Royal Society A: Mathematical, Physical and Engineering Sciences}, 381(2251):20220044, 2023.

\bibitem{dehaene2009origins}
Stanislas Dehaene.
\newblock Origins of mathematical intuitions: The case of arithmetic.
\newblock {\em Annals of the New York Academy of Sciences}, 1156(1):232--259, 2009.

\bibitem{smith2019promise}
Brian~Cantwell Smith.
\newblock {\em The promise of artificial intelligence: Reckoning and judgment}.
\newblock The MIT Press, 2019.

\bibitem{mancosu2001mathematical}
Paolo Mancosu.
\newblock Mathematical explanation: Problems and prospects.
\newblock {\em Topoi}, 20(1):97--117, 2001.

\bibitem{silver1996analysis}
Edward~A. Silver and Jinfa Cai.
\newblock An analysis of arithmetic problem posing by middle school students.
\newblock {\em Journal for research in mathematics education}, 27(5):521--539, 1996.

\bibitem{schulz2012finding}
Laura Schulz.
\newblock Finding new facts; thinking new thoughts.
\newblock {\em Advances in child development and behavior}, 43:269--94, 12 2012.

\bibitem{mcclelland2022capturing}
James~L. McClelland.
\newblock Capturing advanced human cognitive abilities with deep neural networks.
\newblock {\em Trends in Cognitive Sciences}, 26(12):1047--1050, 2022.

\bibitem{carey2009origin}
Susan Carey.
\newblock {\em The Origin of Concepts}.
\newblock Oxford University Press, 2009.

\bibitem{tenenbaum2011how}
Joshua~B. Tenenbaum, Charles Kemp, Thomas~L. Griffiths, and Noah~D. Goodman.
\newblock How to grow a mind: Statistics, structure, and abstraction.
\newblock {\em Science}, 331(6022):1279--1285, 2011.

\bibitem{lake2015human}
Brenden~M. Lake, Ruslan Salakhutdinov, and Joshua~B. Tenenbaum.
\newblock Human-level concept learning through probabilistic program induction.
\newblock {\em Science}, 350(6266):1332--1338, 2015.

\bibitem{gopnik2003theory}
Alison Gopnik.
\newblock The theory theory as an alternative to the innateness hypothesis.
\newblock {\em Chomsky and his critics}, pages 238--254, 2003.

\bibitem{kaminski2008advantage}
Jennifer~A. Kaminski, Vladimir~M. Sloutsky, and Andrew~F. Heckler.
\newblock The advantage of abstract examples in learning math.
\newblock {\em Science}, 320(5875):454--455, 2008.

\bibitem{mitchell2021abstraction}
Melanie Mitchell.
\newblock Abstraction and analogy-making in artificial intelligence.
\newblock {\em Annals of the New York Academy of Sciences}, 1505(1):79--101, 2021.

\bibitem{zhu1987learning}
Xinming Zhu and Herbert~A. Simon.
\newblock Learning mathematics from examples and by doing.
\newblock {\em Cognition and instruction}, 4(3):137--166, 1987.

\bibitem{dehaene1992varieties}
Stanislas Dehaene.
\newblock Varieties of numerical abilities.
\newblock {\em Cognition}, 44(1-2):1--42, 1992.

\bibitem{alcock2015investigating}
Lara Alcock, Mark Hodds, Somali Roy, and Matthew Inglis.
\newblock Investigating and improving undergraduate proof comprehension.
\newblock {\em Notices of the AMS}, 62(7):742--752, 2015.

\bibitem{lester2016can}
Frank~K. Lester and Jinfa Cai.
\newblock Can mathematical problem solving be taught? {Preliminary} answers from 30 years of research.
\newblock In Patricio Felmer, Erkki Pehkonen, and Jeremy Kilpatrick, editors, {\em Posing and Solving Mathematical Problems: Advances and New Perspectives}, pages 117--135. Springer International Publishing, Cham, 2016.

\bibitem{block1986advertisement}
Ned Block.
\newblock Advertisement for a semantics for psychology.
\newblock {\em Midwest Studies in Philosophy}, 10(1), 1986.

\bibitem{murphy2004big}
Gregory Murphy.
\newblock {\em The big book of concepts}.
\newblock The MIT press, 2004.

\bibitem{goodman2014concepts}
Noah~D. Goodman, Joshua~B. Tenenbaum, and Tobias Gerstenberg.
\newblock Concepts in a probabilistic language of thought.
\newblock In Eric Margolis and Stephen Laurence, editors, {\em The Conceptual Mind: New Directions in the Study of Concepts}. The MIT Press, 2015.

\bibitem{lake2023word}
Brenden~M. Lake and Gregory~L. Murphy.
\newblock Word meaning in minds and machines.
\newblock {\em Psychological Review}, 130(2):401–431, 2023.

\bibitem{spelke2000core}
Elizabeth~S. Spelke.
\newblock Core knowledge.
\newblock {\em American psychologist}, 55(11):1233, 2000.

\bibitem{spelke2007core}
Elizabeth~S. Spelke and Katherine~D. Kinzler.
\newblock Core knowledge.
\newblock {\em Developmental Science}, 10(1):89--96, 2007.

\bibitem{dehaene2011number}
Stanislas Dehaene.
\newblock {\em The number sense: How the mind creates mathematics}.
\newblock Oxford University Press, 2011.

\bibitem{feigenson2004core}
Lisa Feigenson, Stanislas Dehaene, and Elizabeth Spelke.
\newblock Core systems of number.
\newblock {\em Trends in cognitive sciences}, 8(7):307--314, 2004.

\bibitem{margolis2015the}
Eric Margolis and Stephen Laurence.
\newblock {\em The Conceptual Mind: New Directions in the Study of Concepts}.
\newblock The MIT Press, 2015.

\bibitem{fodor1975language}
Jerry~A Fodor.
\newblock {\em The language of thought}, volume~5.
\newblock Harvard university press, 1975.

\bibitem{piantadosi2016logical}
Steven~T. Piantadosi, Joshua~B. Tenenbaum, and Noah~D. Goodman.
\newblock The logical primitives of thought: {Empirical} foundations for compositional cognitive models.
\newblock {\em Psychological Review}, 123(4):392--424, 2016.

\bibitem{dehaene2022symbols}
Stanislas Dehaene, Fosca {Al Roumi}, Yair Lakretz, Samuel Planton, and Mathias Sablé-Meyer.
\newblock Symbols and mental programs: a hypothesis about human singularity.
\newblock {\em Trends in Cognitive Sciences}, 26(9):751--766, 2022.

\bibitem{wong2023word}
Lionel Wong, Gabriel Grand, Alexander~K. Lew, Noah~D. Goodman, Vikash~K. Mansinghka, Jacob Andreas, and Joshua~B. Tenenbaum.
\newblock From word models to world models: Translating from natural language to the probabilistic language of thought.
\newblock {\em arXiv preprint arXiv:2306.12672}, 2023.

\bibitem{sable-meyer2022language}
Mathias Sabl{\'e}-Meyer, Kevin Ellis, Josh Tenenbaum, and Stanislas Dehaene.
\newblock A language of thought for the mental representation of geometric shapes.
\newblock {\em Cognitive Psychology}, 139:101527, 2022.

\bibitem{li2022emergent}
Kenneth Li, Aspen~K. Hopkins, David Bau, Fernanda Vi{\'e}gas, Hanspeter Pfister, and Martin Wattenberg.
\newblock Emergent world representations: Exploring a sequence model trained on a synthetic task.
\newblock {\em arXiv preprint arXiv:2210.13382}, 2022.

\bibitem{matsuo2022deep}
Yutaka Matsuo, Yann LeCun, Maneesh Sahani, Doina Precup, David Silver, Masashi Sugiyama, Eiji Uchibe, and Jun Morimoto.
\newblock Deep learning, reinforcement learning, and world models.
\newblock {\em Neural Networks}, 152:267--275, 2022.

\bibitem{mialon2023augmented}
Gr{\'e}goire Mialon, Roberto Dess{\`\i}, Maria Lomeli, Christoforos Nalmpantis, Ram Pasunuru, Roberta Raileanu, Baptiste Rozi{\`e}re, Timo Schick, Jane Dwivedi-Yu, Asli Celikyilmaz, et~al.
\newblock Augmented language models: A survey.
\newblock {\em arXiv preprint arXiv:2302.07842}, 2023.

\bibitem{tenenbaum2006theory}
Joshua~B. Tenenbaum, Thomas~L. Griffiths, and Charles Kemp.
\newblock Theory-based bayesian models of inductive learning and reasoning.
\newblock {\em Trends in cognitive sciences}, 10(7):309--318, 2006.

\bibitem{ullman2020bayesian}
Tomer~D. Ullman and Joshua~B. Tenenbaum.
\newblock Bayesian models of conceptual development: Learning as building models of the world.
\newblock {\em Annual Review of Developmental Psychology}, 2:533--558, 2020.

\bibitem{johnson-laird1989mental}
Philip~N. Johnson-Laird.
\newblock {\em Mental models}.
\newblock The MIT Press, 1989.

\bibitem{rule2020child}
Joshua~S. Rule, Joshua~B. Tenenbaum, and Steven~T. Piantadosi.
\newblock The child as hacker.
\newblock {\em Trends in cognitive sciences}, 24(11):900--915, 2020.

\bibitem{lake2017building}
Brenden~M. Lake, Tomer~D. Ullman, Joshua~B. Tenenbaum, and Samuel~J. Gershman.
\newblock Building machines that learn and think like people.
\newblock {\em Behavioral and Brain Sciences}, 40:e253, 2017.

\bibitem{pearl2018book}
Judea Pearl and Dana Mackenzie.
\newblock {\em The book of why: The new science of cause and effect}.
\newblock Basic books, 2018.

\bibitem{quine1951two}
W.~V.~O. Quine.
\newblock Two dogmas of empiricism.
\newblock {\em The Philosophical Review}, 60(1):20--43, 1951.

\bibitem{aaronson2016pnp}
Scott Aaronson.
\newblock $\mathsf{P}\stackrel{?}{=}\mathsf{NP}$.
\newblock In John~Forbes Nash, Jr. and Michael~Th. Rassias, editors, {\em Open Problems in Mathematics}, pages 1--122. Springer International Publishing, Cham, 2016.

\bibitem{gowers2023mathbelief}
Timothy Gowers.
\newblock What makes mathematicians believe unproved mathematical statements?
\newblock {\em Annals of Mathematics and Philosophy}, 1(1), 2023.

\bibitem{radford2019language}
Alec Radford, Jeff Wu, Rewon Child, David Luan, Dario Amodei, and Ilya Sutskever.
\newblock Language models are unsupervised multitask learners, 2019.

\bibitem{bratman1987intention}
Michael Bratman.
\newblock {\em Intention, plans, and practical reason}.
\newblock CSLI Publications, 1987.

\bibitem{tomasello2022evolution}
Michael Tomasello.
\newblock {\em The evolution of agency: Behavioral organization from lizards to humans}.
\newblock MIT Press, 2022.

\bibitem{newell1972human}
Allen Newell and Herbert~A. Simon.
\newblock {\em Human problem solving}.
\newblock Prentice-Hall, 1972.

\bibitem{tomov2020discovery}
Momchil~S. Tomov, Samyukta Yagati, Agni Kumar, Wanqian Yang, and Samuel~J. Gershman.
\newblock Discovery of hierarchical representations for efficient planning.
\newblock {\em PLoS computational biology}, 16(4):e1007594, 2020.

\bibitem{ho2022people}
Mark~K. Ho, David Abel, Carlos~G. Correa, Michael~L. Littman, Jonathan~D. Cohen, and Thomas~L. Griffiths.
\newblock People construct simplified mental representations to plan.
\newblock {\em Nature}, 606(7912):129--136, 2022.

\bibitem{ho2022planning}
Mark~K. Ho, Rebecca Saxe, and Fiery Cushman.
\newblock Planning with theory of mind.
\newblock {\em Trends in Cognitive Sciences}, 26(11):959--971, 2022.

\bibitem{correa2023humans}
Carlos~G. Correa, Mark~K. Ho, Frederick Callaway, Nathaniel~D. Daw, and Thomas~L. Griffiths.
\newblock Humans decompose tasks by trading off utility and computational cost.
\newblock {\em PLOS Computational Biology}, 19(6):e1011087, 2023.

\bibitem{adamson1952functional}
Robert~E Adamson.
\newblock Functional fixedness as related to problem solving: a repetition of three experiments.
\newblock {\em Journal of experimental psychology}, 44(4):288, 1952.

\bibitem{ho2023rational}
Mark~K. Ho, Jonathan~D. Cohen, and Thomas~L. Griffiths.
\newblock Rational simplification and rigidity in human planning.
\newblock {\em PsyArXiv}, 2023.

\bibitem{arkes1985psychology}
Hal~R. Arkes and Catherine Blumer.
\newblock The psychology of sunk cost.
\newblock {\em Organizational behavior and human decision processes}, 35(1):124--140, 1985.

\bibitem{shafir2002rationality}
Eldar Shafir and Robyn~A. LeBoeuf.
\newblock Rationality.
\newblock {\em Annual Review of Psychology}, 53(1):491--517, 2002.

\bibitem{gershman2015computational}
Samuel~J. Gershman, Eric~J. Horvitz, and Joshua~B. Tenenbaum.
\newblock Computational rationality: A converging paradigm for intelligence in brains, minds, and machines.
\newblock {\em Science}, 349:273--278, 2015.

\bibitem{lieder2020resource}
Falk Lieder and Thomas~L. Griffiths.
\newblock Resource-rational analysis: Understanding human cognition as the optimal use of limited computational resources.
\newblock {\em Behavioral and Brain Sciences}, 43:e1, 2020.

\bibitem{griffiths2020understanding}
Thomas~L. Griffiths.
\newblock Understanding human intelligence through human limitations.
\newblock {\em Trends in Cognitive Sciences}, 24(11):873--883, 2020.

\bibitem{icard2023resource}
Thomas Icard.
\newblock Resource rationality.
\newblock Book manuscript, 2023.

\bibitem{hoffmann2022training}
Jordan Hoffmann, Sebastian Borgeaud, Arthur Mensch, Elena Buchatskaya, Trevor Cai, Eliza Rutherford, Diego de~Las Casas, Lisa~Anne Hendricks, Johannes Welbl, Aidan Clark, et~al.
\newblock Training compute-optimal large language models.
\newblock {\em arXiv preprint arXiv:2203.15556}, 2022.

\bibitem{schwartz2020green}
Roy Schwartz, Jesse Dodge, Noah~A. Smith, and Oren Etzioni.
\newblock Green {AI}.
\newblock {\em Commun. ACM}, 63(12):54–63, 2020.

\bibitem{lombrozo2006structure}
Tania Lombrozo.
\newblock The structure and function of explanations.
\newblock {\em Trends in cognitive sciences}, 10(10):464--470, 2006.

\bibitem{chi1989self}
Michelene~T.H. Chi, Miriam Bassok, Matthew~W Lewis, Peter Reimann, and Robert Glaser.
\newblock Self-explanations: How students study and use examples in learning to solve problems.
\newblock {\em Cognitive science}, 13(2):145--182, 1989.

\bibitem{chi1994eliciting}
Michelene~T.H. Chi, Nicholas De~Leeuw, Mei-Hung Chiu, and Christian LaVancher.
\newblock Eliciting self-explanations improves understanding.
\newblock {\em Cognitive science}, 18(3):439--477, 1994.

\bibitem{williams2010role}
Joseph~J. Williams and Tania Lombrozo.
\newblock The role of explanation in discovery and generalization: Evidence from category learning.
\newblock {\em Cognitive science}, 34(5):776--806, 2010.

\bibitem{bonawitz2012children}
Elizabeth~Baraff Bonawitz, Tessa~J.P. {van Schijndel}, Daniel Friel, and Laura Schulz.
\newblock Children balance theories and evidence in exploration, explanation, and learning.
\newblock {\em Cognitive Psychology}, 64(4):215--234, 2012.

\bibitem{hodds2014self}
Mark Hodds, Lara Alcock, and Matthew Inglis.
\newblock Self-explanation training improves proof comprehension.
\newblock {\em Journal for Research in Mathematics Education}, 45(1):62--101, 2014.

\bibitem{rittle-johnson2017developing}
Bethany Rittle-Johnson.
\newblock Developing mathematics knowledge.
\newblock {\em Child Development Perspectives}, 11(3):184--190, 2017.

\bibitem{wei2022chain}
Jason Wei, Xuezhi Wang, Dale Schuurmans, Maarten Bosma, Fei Xia, Ed~H. Chi, Quoc~V. Le, Denny Zhou, et~al.
\newblock Chain-of-thought prompting elicits reasoning in large language models.
\newblock {\em Advances in Neural Information Processing Systems}, 35:24824--24837, 2022.

\bibitem{zelikman2022star}
Eric Zelikman, Yuhuai Wu, Jesse Mu, and Noah Goodman.
\newblock Star: Bootstrapping reasoning with reasoning.
\newblock {\em Advances in Neural Information Processing Systems}, 35:15476--15488, 2022.

\bibitem{shinn2023reflexion}
Noah Shinn, Federico Cassano, Beck Labash, Ashwin Gopinath, Karthik Narasimhan, and Shunyu Yao.
\newblock Reflexion: Language agents with verbal reinforcement learning.
\newblock {\em arXiv preprint arXiv:2303.11366}, 2023.

\bibitem{lombrozo2011instrumental}
Tania Lombrozo.
\newblock The instrumental value of explanations.
\newblock {\em Philosophy Compass}, 6(8):539--551, 2011.

\bibitem{grice1975logic}
Herbert~P. Grice.
\newblock Logic and conversation.
\newblock In {\em Speech acts}, pages 41--58. Brill, 1975.

\bibitem{levinson1983pragmatics}
Stephen~C. Levinson.
\newblock {\em Pragmatics}.
\newblock Cambridge Textbooks in Linguistics. Cambridge University Press, 1983.

\bibitem{sperber1995relevance}
Dan Sperber and Wilson Deirdre.
\newblock {\em Relevance: Communication and cognition}.
\newblock Blackwell Publishing, 2 edition, 1995.

\bibitem{goodman2016pragmatic}
Noah~D. Goodman and Michael~C. Frank.
\newblock Pragmatic language interpretation as probabilistic inference.
\newblock {\em Trends in cognitive sciences}, 20(11):818--829, 2016.

\bibitem{paolacci2010running}
Gabriele Paolacci, Jesse Chandler, and Panagiotis~G. Ipeirotis.
\newblock Running experiments on {Amazon Mechanical Turk}.
\newblock {\em Judgment and Decision making}, 5(5):411--419, 2010.

\bibitem{palan2018prolific}
Stefan Palan and Christian Schitter.
\newblock Prolific.ac---{A} subject pool for online experiments.
\newblock {\em Journal of Behavioral and Experimental Finance}, 17:22--27, 2018.

\bibitem{sherif1988robbers}
Muzafer Sherif, O.J. Harvey, B.~Jack White, William~R. Hood, and Carolyn~W. Sherif.
\newblock {\em The Robbers Cave experiment: Intergroup conflict and cooperation}.
\newblock Wesleyan University Press, 1988.

\bibitem{campbell2002deep}
Murray Campbell, A~Joseph Hoane~Jr., and Feng-hsiung Hsu.
\newblock Deep {Blue}.
\newblock {\em Artificial intelligence}, 134(1-2):57--83, 2002.

\bibitem{silver2016mastering}
David Silver, Aja Huang, Chris~J. Maddison, Arthur Guez, Laurent Sifre, George van~den Driessche, Julian Schrittwieser, Ioannis Antonoglou, Veda Panneershelvam, Marc Lanctot, Sander Dieleman, Dominik Grewe, John Nham, Nal Kalchbrenner, Ilya Sutskever, Timothy Lillicrap, Madeleine Leach, Koray Kavukcuoglu, Thore Graepel, and Demis Hassabis.
\newblock Mastering the game of {Go} with deep neural networks and tree search.
\newblock {\em Nature}, 529(7587):484--489, 2016.

\bibitem{meta2022human}
Meta Fundamental AI Research Diplomacy~Team (FAIR)†, Anton Bakhtin, Noam Brown, Emily Dinan, Gabriele Farina, Colin Flaherty, Daniel Fried, Andrew Goff, Jonathan Gray, Hengyuan Hu, et~al.
\newblock Human-level play in the game of diplomacy by combining language models with strategic reasoning.
\newblock {\em Science}, 378(6624):1067--1074, 2022.

\bibitem{gobet2004moves}
Fernand Gobet, Jean Retschitzki, and Alex de~Voogt.
\newblock {\em Moves in mind: The psychology of board games}.
\newblock Psychology Press, 2004.

\bibitem{tsividis2017human}
Pedro~A. Tsividis, Thomas Pouncy, Jaqueline~L. Xu, Joshua~B. Tenenbaum, and Samuel~J. Gershman.
\newblock Human learning in {Atari}.
\newblock In {\em 2017 AAAI Spring Symposium Series}, 2017.

\bibitem{allen2023using}
Kelsey Allen, Franziska Br\"{a}ndle, Matthew Botvinick, Judith~E. Fan, Samuel~J. Gershman, Alison Gopnik, Thomas~L. Griffiths, Joshua~K. Hartshorne, Tobias~U. Hauser, Mark~K. Ho, Joshua~R. de~Leeuw, Wei~Ji Ma, Kou Murayama, Jonathan~D. Nelson, Bas van Opheusden, Thomas Pouncy, Janet Rafner, Iyad Rahwan, Robb Rutledge, Jacob~Friis Sherson, Ozgur Simsek, Hugo Spiers, Christopher Summerfield, Mirko Thalmann, Natalia V{\'{e}}lez, Andrew~J. Watrous, Joshua~B. Tenenbaum, and Eric Schulz.
\newblock Using games to understand the mind.
\newblock {\em PsyArXiv}, 2023.

\bibitem{poesia2023research}
Gabriel Poesia.
\newblock Research agenda.
\newblock https://gpoesia.com/research/.
\newblock Accessed: 2023-10-02.

\end{thebibliography}

\end{document}